\documentclass[twocolumn]{article}
\usepackage[T1]{fontenc}

\usepackage{hyperref}
\usepackage{graphicx}
\usepackage{xcolor}
\usepackage{amsmath}
\usepackage{amssymb}
\usepackage{booktabs}
\usepackage[numbers,sort&compress]{natbib}
\usepackage{appendix}
\usepackage{xurl}
\hypersetup{colorlinks=true, linkcolor=[HTML]{1F4E79}, citecolor=[HTML]{1F4E79}, urlcolor=[HTML]{1F4E79}}

\usepackage{fontawesome5}
\usepackage{enumitem}
\usepackage{placeins}

\definecolor{ctrlblue}{HTML}{3498DB} % house style: control
\definecolor{trtred}{HTML}{E74C3C}   % house style: treatment

\newcommand{\highlighttag}[1]{{\textit{#1}}}
\newcommand{\Description}[1]{}  % acmart alt text; not rendered

\usepackage[letterpaper,
            includeheadfoot,
            layoutsize={8.125in,10.875in},
            layouthoffset=0.1875in,
            layoutvoffset=0.0625in,
            left=38.5pt,
            right=43pt,
            top=43pt,
            bottom=44pt,
            headheight=0pt,
            headsep=-2pt,
            columnsep=0.3125in,
            footskip=25pt,
            marginparwidth=38pt]{geometry}

\usepackage{authblk}

\usepackage{caption}
\title{\LARGE Beyond the Townhall: An Integrated Information Environment for Scalable Participatory Urban Planning}
\date{\today}
\author[1]{Carina I. Hausladen*}
\author[2]{Javier Argota Sánchez-Vaquerizo}
\author[2]{Michael Siebenmann}
\author[2]{Arthur Capozzi}
\author[2]{Sachit Mahajan}
\author[2]{Dirk Helbing}

\affil[1]{University of Konstanz}
\affil[2]{Computational Social Science, ETH Zurich}
\affil[ ]{carina.hausladen@uni-konstanz.de; \{javier.argota, michael.siebenmann, arthur.capozzi, sachit.mahajan, dirk.helbing\}@gess.ethz.ch}

\begin{document}
\twocolumn[
  \begin{@twocolumnfalse}
\maketitle

\begin{abstract}
\noindent
LLMs are increasingly explored for public participation, but less is known about how they should be designed around the information citizens encounter beforehand.
In an experiment (N=200) on participatory urban planning, participants learned about a sustainability project adapted from a real street redesign, through text and images or narrated 360-degree video with spatial audio and spatial anchoring; they subsequently engaged with factual and reflective LLM personas.
The immersive condition improved recall and changed what participants said to the agents: they named beneficiaries more often, asked directionally more project-related questions, and provided more positive feedback.
We contribute 
(1) a conceptualization of LLMs as dialogue partners complementing a shared information baseline,
(2) causal evidence that this baseline shapes those conversations,
(3) an account of how citizens use and judge factual versus reflective agent roles, and
(4) a method for cities to test information packages and agent configurations before public release.
\vspace{0.50cm}
\end{abstract}
  \end{@twocolumnfalse}
]

\renewcommand{\thefootnote}{\fnsymbol{footnote}}
\footnotetext[0]{*Corresponding Author: carina.hausladen@uni-konstanz.de}
\renewcommand{\thefootnote}{\arabic{footnote}}

\section{Introduction}
In participatory urban planning, citizens are increasingly asked to make consequential decisions about technically complex projects. 
In November 2016, for example, voters in Los Angeles County approved Measure M, a permanent sales tax expected to raise more than \$120 billion over four decades of transit expansion \cite{lametro2016measurem,manville2018measurem}. 
On the same day, voters in the Seattle region approved Sound Transit 3 (ST3), a \$53.9 billion plan to more than double the region's light-rail system \cite{soundtransit2016st3,soundtransit2016systemplan}. 
Decisions of this kind require citizens to make judgments about information that may be unfamiliar, technical, and consequential. 
Yet participatory processes must enable all citizens, not only technical experts, to acquire and make sense of this information.

Deliberative democracy offers one established response. In a typical deliberative process, participants first receive expert-curated briefing materials that establish a shared informational baseline and then deliberate in small groups before forming a collective judgment \cite{fishkin2018democracy}.

\begin{figure*}[t]
\centering
  \includegraphics[width=\textwidth]{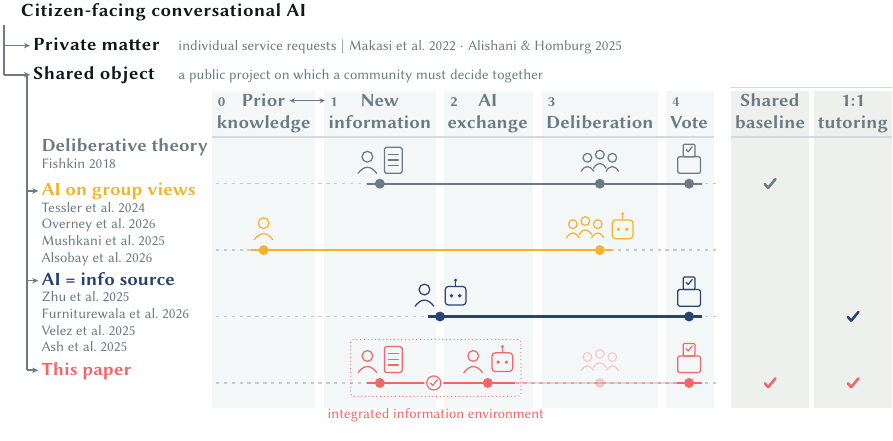}
\caption{{\bf Position of AI in the deliberation process: the missing combination of shared information and individual dialogue.}
Theory (top row) asks citizens to read a shared, expert-curated baseline, deliberate in small groups, and then vote. One strand of AI research (gold) places the agent within the deliberation stage, typically on topics participants already know about. 
A second strand (navy) uses dialogue as the primary mechanism for acquiring information, replacing the shared baseline. 
The two rightmost columns capture two requirements: a baseline that provides the same core information to every citizen, and a counterpart that can respond to each citizen's individual questions. 
Our approach (coral) combines both: curated information about a project genuinely new to participants, a check of what they retained, and a one-to-one AI dialogue partner that helps them make sense of that information before deliberation or voting.}
  \label{fig:spine}
  \Description{A matrix with one row per research line and one column per stage of the deliberative process, from information provision to vote. Icons in each cell show what the citizen has at that stage: a booklet, a chatbot, or neither. Two rightmost columns mark whether the line supplies a curated baseline identical for every citizen and a counterpart answering the citizen's own questions. Only the bottom row, this paper's design, is marked in both.}
\end{figure*}

With the increasing capabilities of general-purpose LLMs, HCI research has begun to explore how AI might support participatory and deliberative processes. Two approaches are particularly relevant. 
The first places LLMs within the deliberation stage itself (\autoref{fig:spine}, gold row). For example, researchers have examined how AI can mediate participants' private opinions into a group statement \cite{tessler2024ai}, or synthesize thousands of public comments \cite{overney2026narrative}. 
The second uses an LLM as the primary interface through which citizens acquire information (\autoref{fig:spine}, navy row). 
For example by answering questions about ballot measures \cite{ash2025ballotbot}, providing voting advice grounded in party platforms \cite{velez2025chatbot,zhu2025vaa}, or engaging citizens in conversations about topics such as crime, air quality, and unemployment \cite{furniturewala2026dialogue}.

These approaches address different needs, but neither addresses the combination that real-world participatory urban planning requires. 
The experiments on deliberation-support systems typically ask participants to deliberate about topics on which they already possess some knowledge. This may be partly a methodological necessity: it is difficult to elicit genuine discussion in a scientific experiment when participants are asked to deliberate about a topic they know little about. The Habermas Machine, for example, studied deliberation about immigration, Brexit, and the minimum wage \cite{tessler2024ai}---topics that participants are likely to have encountered before and that therefore make it possible to generate genuine discussion in a laboratory setting. 
This line of research consequently begins after citizens have already acquired the knowledge needed for deliberation.

Information-oriented systems address the earlier stage, but make dialogue itself the primary source of information. In doing so, they replace a shared, curated informational baseline. 
Dialogue, however, is a poor substitute for curation.
Learners do not know what they do not understand, and frequently overestimate the depth of their understanding \cite{rozenblit2002illusion}. Nor do group settings encourage extensive questioning: students ask substantially more questions when receiving one-to-one tutoring than in conventional classroom instruction \cite{graesser1994question}. Curated material, in contrast, can establish a \emph{common starting point} and ensure that important information is presented regardless of whether a citizen thinks to ask for it.
This is reflected in established democratic practice. Before federal votes, for example, the Swiss Federal Council sends the \emph{Abstimmungsbüchlein} (explanatory booklet) to households, providing voters with curated information about the proposals and the arguments surrounding them \cite{swiss2021erlaeuterungen}.

The practical challenge is therefore not to add AI to deliberation, or to replace briefing materials with a chatbot. It is to determine whether an AI dialogue partner can operate \emph{between} a shared informational baseline and subsequent deliberation or voting.

LLMs make such individualized interaction potentially scalable. One-to-one tutoring has long been associated with substantially stronger learning outcomes than conventional classroom instruction; Bloom famously found a two-standard-deviation difference between tutored and conventionally taught students \cite{bloom1984sigma}. 
LLMs create the possibility of providing some of the benefits of one-to-one interaction without requiring a human tutor for every participant. 

This suggests a role for AI that differs from both a dialogue-information source and a deliberation facilitator: an LLM can act as an individualized \emph{sensemaking partner} that sits on top of a shared informational baseline and helps citizens interrogate, clarify, and explore what they have learned (\autoref{fig:spine}, bottom row).

% Two open questions
Once AI occupies this role, two important design questions arise. The first concerns the \emph{normative stance} of the dialogue partner. Common wisdom asks for an unbiased or neutral AI, but neutrality is not straightforward. Habermas \cite{habermas1984theory} argues that arguments are evaluated according to standards of truth, sincerity, and normative rightness. Critical theorists such as Foucault \cite{foucault1980power} emphasize that discourse is embedded within historically structured relations of power. Urban planning has a name for one such historically entrenched structure: moto-normativity, the privilege of automobility in policy-making and public discourse \cite{Walker2023InternationalHazard}. Where the range of acceptable arguments is already structured in this way, simply presenting ``both sides'' does not necessarily produce a neutral interaction \cite{fraser1990rethinking,kelly2023habermasian,hammond2025deliberative}. 
We therefore treat the normative stance of the dialogue partner as an empirical design question and not as a specification that can be settled theoretically in advance.

The second concerns the agent's \emph{conversational behavior}. Citizens may have different needs after encountering the same information baseline. Some may primarily seek factual clarification; others may want to reflect on implications, explore alternatives, or connect the project to their own concerns. An agent optimized for factual precision may provide useful answers while offering little opportunity for reflection. An agent designed to stimulate reflection may support deeper engagement, but may also introduce framing effects of its own. The appropriate balance cannot be determined solely from the capabilities of the underlying model. It depends on how a particular public responds to the agent in the context of a particular project and informational environment.

These observations suggest a broader HCI challenge for civic AI: instead of assuming a universally appropriate AI persona, we need ways to \emph{diagnose} which configurations work for the people and contexts in which they will actually be deployed. 
Recent work in science \cite{ashokkumar2026large,li2026matraix,novelli2026replica} and industry \cite{simile,electrictwin,artificialsocieties} has increasingly explored populations of LLMs as synthetic audiences for addressing such questions. However, critical research has shown that synthetic audiences do not necessarily reproduce the breadth of preferences expressed by real people \cite{yang2024llmvoting}. For participatory systems, where the preferences of the affected public are themselves the object of concern, this limitation makes human-centered evaluation especially important.

We therefore propose a \emph{human-centered route to evaluating civic AI}: an online experimental procedure in which humans can be exposed to a project's information materials and alternative AI dialogue-partner configurations before those systems are deployed in an actual participatory process. 
Such a procedure could allow municipalities to test not only whether an AI system is useful in the abstract, but whether a particular combination of information materials and AI behavior works for the citizens who will ultimately use it.

The central question we ask in this paper is:
\emph{How can AI support participatory deliberative processes by complementing a shared informational baseline?}

Drawing on deliberative-democracy theory, learning research, and prior work on civic LLMs, we identify a role for AI as an individualized \emph{dialogue partner between curated information and subsequent deliberation or voting}. 
This role combines two complementary requirements: citizens should encounter a common set of core information, but they should also have an opportunity to ask questions and make sense of that information in ways that are specific to their own needs.

This framing gives rise to two empirical questions. 
\begin{itemize}
\item First, how does the design of the shared informational baseline shape citizens' subsequent interaction with an AI dialogue partner?
\item Second, how do citizens perceive and use different dialogue-partner roles after receiving that baseline? 
\end{itemize}

These questions are inherently contextual: the appropriate dialogue-partner design may depend on the public, the project, and what citizens actually retain from the information they receive.

We investigate these questions through an online human-subjects experiment situated in an urban-planning scenario. Participants learn about a real-world urban sustainability project involving changes to the old town of Lausanne, represented through a digital twin reconstructed from a high-resolution photogrammetric drone scan. Participants are placed in the role of citizens of a fictitious city and the project, information materials, and omissions are closely mirrored from corresponding field documents.

We vary how participants are informed about the project \emph{between} participants. A control group receives information in a format typical of many public consultations: a webpage containing text and static images. 
A treatment group receives the same substantive facts in an immersive format designed to support recall, combining narrated video, spatial audio, and mnemonic techniques that spatially anchor key facts. After the information phase, we measure what participants retained about the project. 
Because information format is randomized, we can identify its \emph{causal effect} on retention and examine whether differences in retained knowledge subsequently shape participants' interactions with AI dialogue partners.

Each participant then interacts with two LLM personas: first a factual persona and subsequently a reflective persona. This within-participant comparison allows us to examine how citizens perceive and use different dialogue-partner roles in a participatory urban-planning setting, including what they ask, how extensively they engage, and whether the information format they received shapes their subsequent interactions with the agents.

The resulting experimental design allows us to examine the interaction between two layers of an emerging civic information environment: a common, curated baseline and an individualized AI dialogue partner. 
It allows us to ask not only whether citizens find an AI dialogue partner useful, but also whether its usefulness depends on what citizens were told beforehand and what they retained from that information. 

More broadly, our paper treats the experiment itself as a \emph{diagnostic instrument}: municipalities and researchers could replace our project materials, target population, and dialogue-partner specifications with their own and use the same procedure to test alternative configurations before deployment.

Our contributions are fourfold. 
\begin{enumerate}
    \item We conceptualize an underexplored role for LLMs in participatory deliberation: an individualized dialogue partner that complements a shared curated information baseline. 
    \item We provide empirical evidence that the design of a shared information environment shapes both what citizens retain and how they subsequently engage with AI dialogue partners.
    \item We characterize how citizens use and evaluate distinct factual and reflective agent roles in an urban-planning context.
    \item We contribute a reusable human-centered evaluation procedure that allows municipalities to stress-test both their information materials and different LLM dialogue partner roles with human participants.
\end{enumerate}

\begin{figure*}
\centering
%figures/graph_abstract_tikz.tex
  \includegraphics[width=\textwidth]{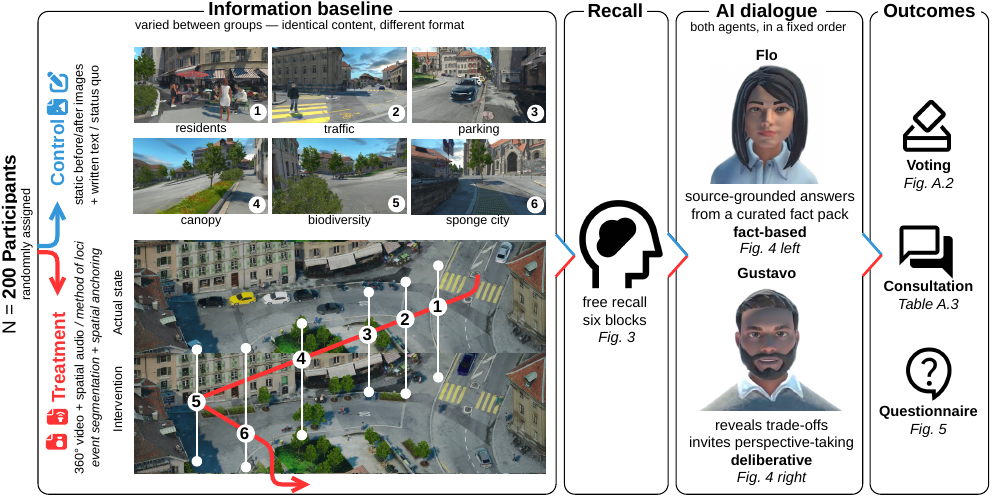}
\caption{{\bf Overview of the behavioral experiment.} The study uses a photogrammetric digital twin of the city center of Lausanne in which an urban sustainability project is implemented. Participants (N=200) were randomly assigned to treatment or control groups and received identical content across six information blocks. The treatment group experienced immersive first-person 360-degree video walkthroughs with spatial audio narration, explicitly designed to support mnemonic encoding. The control group received equivalent information through static images and text in a conventional format. Following information delivery, key outcomes were assessed, including, among others, the interactions with two purpose-built LLM personas: Flo offers source-grounded factual clarifications, and Gustavo facilitates normative reflection, perspective-taking, and deliberative discussion.}
  \label{fig:abstract}
\Description{Flow diagram of the experiment. Two hundred participants are randomly split into a treatment arm receiving narrated 360-degree video walkthroughs and a control arm receiving static images with text, both covering six identical information blocks anchored to locations along a street in a photogrammetric digital twin of Lausanne. Both arms then proceed to a recall task, conversations with two LLM personas named Flo and Gustavo, voting, and consultation feedback.}
\end{figure*}

\section{Related Work} 

\paragraph{Civic technology and participatory urban planning.}
HCI engagement with public participation runs through digital civics, which asks how digital technologies might scaffold a move from transactional to relational models of service provision and governance \cite{vlachokyriakos2016digital}, and through participatory design concern with how publics are constituted rather than merely consulted \cite{ledantec2013infrastructuring}.  Within planning, systems have lowered the cost of contributing: while CommunityCrit decomposes planning documentation into micro-activities, so residents can comment without a large time commitment \cite{mahyar2018communitycrit}, participatory media production has been used to widen participation through storytelling  \cite{Manuel2017}, and digitally supported urban walks have been used to make planning politics tangible \cite{crivellaro2015contesting}. A parallel line studies the municipal side: how officials actually run engagement \cite{corbett2018problem}, and how trust between residents and institutions is built or lost through civic technology \cite{corbett2021trust}. Critical work cautions against ``participation washing'' and extractive forms of community involvement \cite{sloane2022participation}, notes that the design workshop is itself a privileged format that can neglect underserved populations \cite{harrington2019deconstructing}, and shows that participation imposes epistemic burdens on those least able to bear them \cite{pierre2021getting}. Closest to the diagnostic use we propose, CommunityClick captures town-hall feedback and reports it back in a form that planners can act on \cite{jasim2021communityclick}. What this literature has largely left untouched, however, is the information that citizens are provided \emph{before} they contribute.

\paragraph{Immersive media for planning participation.}
Immersive representation has been studied in HCI for over a decade, from smartphone Augmented Reality overlaid on planning sites \cite{allen2011smartphone} to Virtual Reality (VR) used to provoke civic engagement with climate futures \cite{ferris2020melbourne}. Closest to our treatment, van Leeuwen et al.\ found that immersive display raised engagement and the amount citizens recalled of a park design, but changed neither the accuracy of that recall, nor their confidence in the decision, nor the outcome of the municipal ballot in which 1302 residents voted \cite{vanleeuwen2018effectiveness}. This motivates our own constraint: we ask what a briefing format achieves at deployable scale, i.e., using browser-delivered 360-degree video rather than headsets, which a municipality cannot distribute at consultation scale. Our contribution is how the form of that briefing shapes the AI conversation following it.

\paragraph{Conversational systems for deliberation.}
Before LLMs, Computer-Supported Cooperative Work (CSCW) built structured deliberation interfaces: ConsiderIt organised a voters' guide around pro and con trade-offs \cite{kriplean2012considerit}, Reflect asked readers to restate what others meant \cite{kriplean2012reflect}, and Opinion Space mapped distribution of participants' opinions \cite{faridani2010opinion}. These systems embedded deliberative practices in the interface by structuring activities such as considering trade-offs, listening, and attending to diverse perspectives; more recent AI-mediated systems instead use generative models to perform parts of this work, e.g., by synthesising participants' views into statements of common ground \cite{tessler2024ai}. However, recent simulation work suggests that LLM groups reach near-human procedural quality while exhibiting less perspective heterogeneity \cite{flechtner2026procedural}. Gustavo's instruction to reflect participants' input echoes Reflect's design, applied to the participant's own position rather than that of another citizen. How citizens should calibrate trust in such an agent is itself an HCI question: appropriate reliance, rather than high trust, is the design target \cite{lee2004trust}, and cognitive forcing can reduce overreliance on a confident system, though users like it less \cite{bucinca2021trust}. What differs here is that the agent sits after curated information material whose form we manipulate, and that we measure what each participant retained before the exchange begins.

\section{Experimental Design}
In our experiment, participants ($N$ = 200) were asked to take the role of citizens in a fictional city and vote on a proposed street redesign project---a decision of the kind that participatory urban planning addresses. 
The project was inspired by a real redesign project to take place in Lausanne\footnote{\url{https://web.archive.org/web/20250909171826/https://www.lausanne.ch/vie-pratique/mobilite/espaces-publics/avenue-echallens.html}}, Switzerland, but not identical (\ref{apx:vignette}).
It centered on promoting the sustainable and democratic use of public space by adding bike lanes, reducing parking spaces, and introducing benches, greenery, and bike racks.

We visualized the planned changes using a digital twin of the city center of Lausanne, reconstructed from a high-resolution photogrammetric drone scan (\ref{apx:video}). The proposed street modifications were then implemented within this urban digital twin.

We used a between-subjects design to experimentally vary how information was presented to participants. Both groups received identical content across six information blocks: residents, traffic, parking, tree canopy, biodiversity, and sponge pavement (\autoref{fig:abstract} shows the spatial anchoring of these information blocks in the digital twin).

The form of that presentation was designed to maximize recall:
Recall is better when information arrives through verbal \emph{and} visual channels simultaneously \cite{paivio1990mental},
and improves when information is encoded in a structured context and retrieved in a matching one \cite{godden1975context}. The oldest applied form of that last principle is the method of loci, in which facts are placed along a familiar route and recovered by walking it \cite{paivio1990mental,legge2012use}.

\begin{description}[nosep]
\item[{\faFileVideo[regular] \faFileAudio[regular]}] In the {\em treatment} group, participants viewed short 360-degree video clips from a first-person perspective with spatial audio.
The narration explicitly anchored key information to visible landmarks (``method of loci") and supported the imagination of the project implementation proposal. The first-person perspective transitioned sequentially between six locations along the street to create event segmentation boundaries that reinforced the impression of walking forward and strengthened memory encoding through contextual shifts.
\item[{\faFileImage[regular] \faEdit[regular]}]
In the {\em control} group, participants were shown static images of the status quo and the envisioned redesign, with no video. Information was presented as written text, not through narration, and did not explicitly anchor key points to local visual elements.
\end{description}

\noindent We measured several outcome variables:

\begin{description}[nosep]
\item[Recall.] After receiving the information, participants were asked to write down everything they remembered in an open-text field.
\item[LLMs.] Participants engaged with two LLMs, explicitly presented as AI systems. Both drew on the same shared content: a fact pack authored from the City of Lausanne's planning documentation, together with prepared responses to questions we anticipated would be common, including cost and construction timeline. ``Flo'' answered from it, with instructions to prioritise it and flag gaps (\autoref{apx:llm}; prompts released in full in our public repository); ``Gustavo'' facilitated deliberative reflection by revealing trade-offs and encouraging perspective-taking.
\item[Voting.] Participants proceeded to vote on the street redesign project in three ways: (i) approval voting, (ii) explicit ranking, and (iii) giving an overall yes/no vote.
\item[Consultation.] Participants could, via a free-text field, submit consultation proposals to the city.
\item[Questions.] Finally, participants completed questionnaires on their evaluation of the effectiveness of video-based information provision and their traffic habits.
\end{description}

\begin{figure*}[ht]
    \centering
    \includegraphics[width=0.99\linewidth]{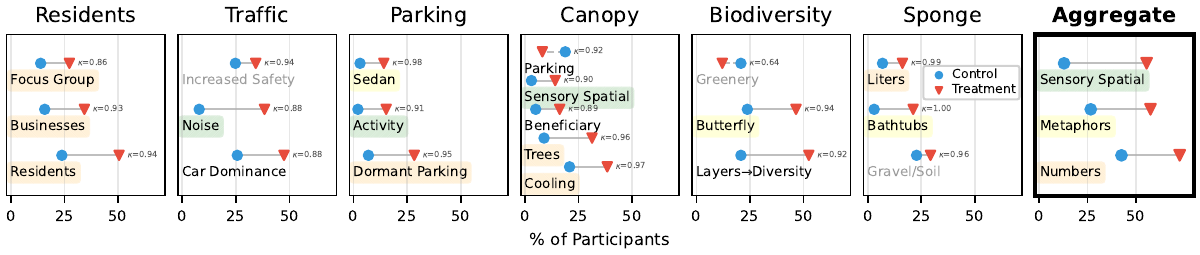}
    \caption{{\bf Memory Task.} Comparison of theme frequencies between the control and treatment groups, for each information block (panels 1-6), and aggregated high-level characteristics (rightmost panel). Each subplot shows only those themes that are recalled by at least ten participants in either the control or treatment group. Labels set in black show significant differences according to a Fisher's exact test ($p<0.05$); labels greyed back are not significant. Color coding corresponds to the three categories shown in the ``Aggregate" panel. Counts are the human-validated annotations. The value beside each theme is Cohen's $\kappa$ for that theme, averaged over the coder pairings its block supports (Appendix~\ref{apx:qual}).}
    \label{fig:memory}
    \Description{Seven panels of horizontal bars comparing how often each recall theme appeared in the control and treatment groups. Panels one to six correspond to the information blocks ``residents'', ``traffic'', ``parking'', ``tree canopy'', ``biodiversity'', and ``sponge pavement''; the rightmost panel aggregates themes into numerical, metaphorical, and sensory-spatial categories. Treatment bars exceed control bars on most themes, with significant differences labelled in black and non-significant ones greyed.}
\end{figure*}

\paragraph{Experiment Ethics.}
The experimental protocol was reviewed and approved by the Ethics Commission of ETH Zurich (EK 2024-N-99).
The experiment was conducted online with participants recruited via Prolific.
Before participation, all subjects provided informed consent electronically.
Participants were informed about the purpose and procedures of the study, their right to withdraw at any time without penalty, and the handling of their data.
The use of LLMs within the experimental design was explicitly reviewed and approved by the Ethics Commission. Participants were clearly informed that they would interact with LLMs during the study and were instructed not to submit personal data such as names or other identifying information in their interactions.
To ensure data protection, experimental response data were stored separately from sociodemographic information. Sociodemographic data were obtained from Prolific and matched via Prolific ID, but were never stored together with experimental response data. The experiment was hosted on a server located in Switzerland.

\section{Results}
The sample included 200 participants recruited via Prolific with a mean age of 40.2 $\pm$ 12.7 years. Participants were mostly residing in the United Kingdom (64.5\%), and were predominantly White (85.0\%), male (57.0\%), and full-time employed (38.0\%, Appendix \autoref{tab:sociodemographics}).

The treatment  ($N = 99$) and control group ($N = 101$) answered control questions in a similar way (\autoref{tab:mobility}): When asked about their mobility habits, most participants stated travelling by car (46.0\%), followed by walking (24.0\%), public transportation (20.0\%), and cycling (10.0\%). The largest group considered neighborhood community to be somewhat important (46.0\%). Neighborhood attractiveness was rated by most respondents as very important (74.5\%).

\paragraph{Audiovisual treatment improved recall quality.}

The treatment condition \faFileVideo[regular]~\faFileAudio[regular] significantly improved the quantity and quality of recalled information compared to the control condition \faFileImage[regular]~\faEdit[regular].

%%%%% QUANTITATIVE ASSESSMENT
We found that treatment participants wrote significantly longer recalls ($\mu$ = 22.3 words) compared to control participants ($\mu$ = 15.9 words; $p<.01$, Cohen's $d=0.58$).

%%%% QUALITY
Additionally, the content of the recalled information was different.
Specifically, we identified three notable qualitative differences in recall characteristics (\autoref{fig:memory}, Panel Aggregate):
First, participants in the treatment group recalled more {\em numerical} facts accurately.
Second, they were more likely to reproduce the {\em metaphors} used to contextualize technical data (e.g., ``the permeable tiles hold 2{,}000\,L of water, equivalent to 10 bathtubs''). This suggests deeper semantic encoding that made the information %relatable and 
easier to imagine and retain.
Third, recall in the treatment group was more frequently grounded in {\em sensory and spatial} impressions, e.g., reductions in noise or a more peaceful atmosphere.

Specific noteworthy differences emerged in sub-aspects of the city renovation scenario.
For the canopy-tree section (\autoref{fig:memory}, Panel~4), participants in the control condition recalled the \textit{absence} of parking more frequently, whereas treatment-group participants focused less on parking removal and more on environmental benefits.
Similarly, in the biodiversity section (\autoref{fig:memory}, Panel~5), the treatment group recalled the intended causal mechanism more than twice as often—namely, that a \emph{layered} planting structure, with vegetation at different heights, supports a wider range of species—whereas control responses more often registered only generic greenery (e.g., ``more trees or flowers'') without the layering mechanism that drives the ecological benefit.

\paragraph{Participants perceive video-audio as the preferred communication format.}
We asked participants to imagine a referendum in their own city on a similar project, with the website presenting the information either as short videos with audio or as the images and written text that most city websites use, and then to answer ``thinking about how the average citizen in your city would respond''.\footnote{We explicitly used third-person framing because first-person questions can trigger identity concerns and self-presentation biases, whereas third-person formulations reduce ego threat and yield more candid evaluations \cite{benabou2006incentives,davison1983third,paulhus1984two}.}

\begin{figure*}[t]
\newcommand{\panelhead}[2]{\parbox[t][2\baselineskip][t]{\linewidth}{\small\raggedright\textbf{#1}\hspace{0.7em}#2\par}\par}
\begin{minipage}[t]{0.40\textwidth}
\vspace{0pt}
\panelhead{a}{Participants' perceptions of information formats (\%)}
\footnotesize
\setlength{\tabcolsep}{3.5pt}
\newcommand{\tabnote}[1]{}%
% generated by scripts/tab_format_preferences.py
\begin{tabular}{l r | l r}
\toprule
\multicolumn{2}{c}{\textbf{Preferred}} &
\multicolumn{2}{c}{\textbf{More engaging}} \\
\midrule
{\color{trtred}$\blacktriangledown$}\,{\faFileVideo[regular] \faFileAudio[regular]} & 76.5
  & {\color{trtred}$\blacktriangledown$}\,{\faFileVideo[regular] \faFileAudio[regular]} & 75.5 \\
{\color{ctrlblue}\scalebox{1.2}{$\bullet$}}\,{\faFileImage[regular] \faEdit[regular]} & 17.5
  & {\color{ctrlblue}\scalebox{1.2}{$\bullet$}}\,{\faFileImage[regular] \faEdit[regular]} & 10.0 \\
No preference & 6.0
  & About the same & 14.5 \\
\midrule
\multicolumn{2}{c}{\textbf{Understand better}} &
\multicolumn{2}{c}{\textbf{Remember better}} \\
\midrule
{\color{trtred}$\blacktriangledown$}\,{\faFileVideo[regular] \faFileAudio[regular]} & 69.5
  & {\color{trtred}$\blacktriangledown$}\,{\faFileVideo[regular] \faFileAudio[regular]} & 57.0 \\
{\color{ctrlblue}\scalebox{1.2}{$\bullet$}}\,{\faFileImage[regular] \faEdit[regular]} & 16.0
  & {\color{ctrlblue}\scalebox{1.2}{$\bullet$}}\,{\faFileImage[regular] \faEdit[regular]} & 26.0 \\
About the same & 14.5
  & About the same & 17.0 \\
\bottomrule
\end{tabular}
\providecommand{\tabnote}[1]{\vspace{2mm}\begin{minipage}{0.95\columnwidth}\footnotesize\textit{Note.} #1\end{minipage}}
\tabnote{$N = 200$; no missing responses. Information was provided via video and audio ({\color{trtred}$\blacktriangledown$}~treatment) or images and text ({\color{ctrlblue}\scalebox{1.2}{$\bullet$}}~control); the marks label the formats, not the respondents, and every percentage is over all participants of both arms. Participants in both groups tended to prefer the video/audio format. All four distributions differed significantly from chance ($\chi^2$ tests, all with $p < 0.001$).}

\end{minipage}
\hfill
\begin{minipage}[t]{252pt}
\vspace{0pt}
\panelhead{b}{Perceptions of LLMs}
\includegraphics[width=252pt]{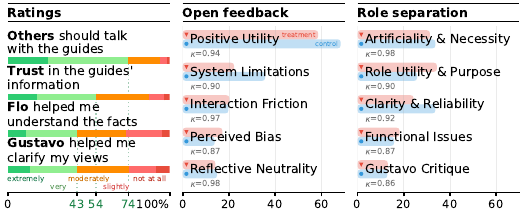}
\end{minipage}
\caption{\textbf{Participant perceptions.} \textbf{(a)} Format preferences: Participants indicated whether people in their city would prefer, better understand, better remember, or be more engaged by video with audio (treatment) or images with text (control). All four distributions differed significantly from chance ($\chi^2$, $p<0.001$). \textbf{(b)} Agent ratings: Participants rated whether others should consult the agents, whether they trusted the agents' statements, and whether Flo aided understanding and Gustavo aided reflection, on a 5-point scale from ``extremely agree'' to ``not at all.''
\emph{Open feedback.} Free-text feedback on the agents yielded five main themes, with no significant differences between treatment and control; Cohen's $\kappa$ is reported for each theme.
\emph{Role separation.} Free-text comments on the distinct roles of Gustavo and Flo yielded five main themes, again with no differences between treatment and control; Cohen's $\kappa$ is reported for each theme.}
\label{fig:guides_rating}
\Description{Two halves. The left half is a table of the percentage of participants who thought the average person in their city would prefer, better understand, better remember, or be more engaged by video with audio versus images with text; video leads on all four. The right half plots participants' ratings of the two agents, their open feedback themes, and their views on keeping the two agent roles separate, with treatment and control shown as paired markers per theme.}
\end{figure*}

76.5\% believed that most citizens would {\em prefer} the video format. 
57.0\% thought that most citizens would {\em remember} the content better using video; 69.5\% thought people would {\em understand} it better via video, and 75.5\% believed that video would be more {\em engaging} ($\chi^2$ tests, all with $p< 0.001$, \autoref{fig:guides_rating}a).
Importantly, the pattern is not an artifact of exposure bias: the treatment group, which had just experienced the video, in fact endorsed the video format \emph{less} than the control group (71\% vs.\ 82\% preferring video; difference $-11$ percentage points, 95\% CI $[-23,\,0]$).
Direct experience tempered rather than inflated enthusiasm for the format, so the overall preference for video does not simply reflect having seen it.

\paragraph{Engagement with LLMs.}
Prior work asked participants to converse with LLMs about topics they already had some knowledge of.
Our experiment, by contrast, introduced participants to new information, altered the provision of this new information (making it easier or harder to retain), and controlled for \textit{retained information} before participants interacted with any LLM.

%% engagement frequency
It is important to highlight that the LLM conversations were not enforced, yet only two participants skipped Flo and one skipped Gustavo (198 and 199 respectively);
Those who engaged had a median of 11 turns and some 177 words with the two agents. Of the 198 participants who wrote to Flo, 197 raised a substantive topic. Participants were, in other words, using the agents to genuinely work through a decision. This speaks to the experimental realism of our study-design and highlights its potential as a diagnostic instrument for municipalities. 

Participants addressed roughly twice as many turns to Gustavo as to Flo (7.8 vs.\ 4.7; Wilcoxon signed-rank $p<0.001$). This directly reflects the two agents' different roles by design: Flo was instructed to ``answer directly and clearly'' and to close each exchange, whereas Gustavo was instructed to ``ask open follow-up questions'' and to ``always reflect the participant's input'' (the complete prompts are released in our public repository; \autoref{apx:llm}).
Because the two agents were presented in a fixed order, this contrast is not experimentally isolated; the most plausible order effect over an eighteen-page study is a declining engagement. 
However, we observe the opposite.
How much a citizen converses with an information agent thus appears to be shaped by the conversational role the agent is given.

Crucially, treatment and control groups did not differ in \emph{how much} they used the agents: the number of turns addressed to Flo was statistically indistinguishable across conditions (4.7 vs.\ 4.7 turns per participant; Mann--Whitney $p=0.91$), as was the number addressed to Gustavo (8.2 vs.\ 7.4; $p=0.20$; 95\% CI on the difference $[-0.7, +2.3]$ turns).
Equal take-up matters for \textit{identification:} because treatment participants did not extract \textit{more} information from the agents than control participants did, differences in what they subsequently recalled and decided cannot be attributed to differential information seeking. They are instead attributable to the audiovisual channel, through which the information was delivered.

\begin{figure*}[t]
  \centering
  \includegraphics[width=\textwidth]{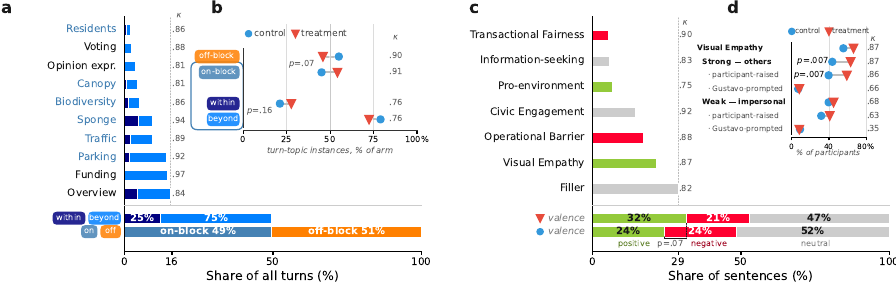}
\caption{\textbf{Conversations with the two agents.} \textbf{(a,b)} Flo: bars show topics, split into questions answerable \emph{within} the shown material (navy) and questions reaching \emph{beyond} it (light blue); blue labels mark the six on-screen information blocks. \textbf{(c,d)} Gustavo: bars show themes, coloured by valence. \textbf{(b,d)} show by-arm contrasts. In (b), turn-topic instances are tested using participant-clustered logistic regression, with one $p$-value per pair of rows. In (c), valence is tested at the sentence level using the same approach. In (d), percentages of participants with at least one such turn are compared using Fisher's exact test; $p$-values are shown only for contrasts reaching $p<0.05$. Cohen's $\kappa$ is reported for every theme (Appendix~\ref{apx:qual}).}
  \label{fig:chats}
  \Description{Four panels. Panels (a) and (b) cover conversations with Flo: horizontal bars give the frequency of each question topic, split by whether the answer lay within or beyond the material shown, with summary bars pooling on-block against off-block topics and by-arm contrasts plotted as paired markers. Panels (c) and (d) cover Gustavo: bars give theme frequency coloured by valence, with a stacked pair showing each arm's valence composition and paired arm markers beneath.}
\end{figure*}

%FLO NEW
We coded the conversations of the 198 participants who wrote to {\bf Flo} along two axes.
The first asks whether a question was \emph{on-block}, that is about one of the six information blocks, or \emph{off-block}. The second applies only to on-block questions and asks whether the answer lay \emph{within} the material shown---e.g. a number the participant had forgotten---or \emph{beyond} it (\autoref{fig:chats}a,\,b).
Most often, participants asked Flo for an \highlighttag{Overview} of the proposal as a whole.
The remaining themes address concrete concerns about what it would mean to live with the redesign. 
\highlighttag{Funding} was the largest of them, with participants asking questions such as, ``Will taxes be increased to pay for this?'' 
That is telling, because the information material, modelled on the official consultation documents, gave neither information on costs nor on timelines.
We knew, however, from pilot runs that participants asked about these issues and therefore explicitly added this information to the LLMs’ shared content. 
%, even though both were included for the agents to be available on request. 
We return to this information gap in the official documents in the discussion.
The treatment condition shifted what participants asked about.
While control participants asked more \emph{off-block} questions (55\%) (e.g. asking about costs, comparisons, procedure and the views they had brought with them), 
treatment participants asked more \emph{on-block} questions (54\%) ($p=0.07$, participant-clustered; Appendix~\ref{apx:qual}).

The direction of the effect suggests that the immersive material gave participants a concrete, shared representation around which to organise their inquiry. In its absence, ``control'' participants appear to have leaned more on the frames they arrived with.
This is in line with research on learning, where the advantage of guidance recedes only once learners hold enough prior knowledge to supply their own \citep{kirschner2006minimal}; unprompted question-asking is correspondingly sparse and shallow \citep{graesser1994question}.

The second axis points the same way. Across both arms, 75\% of on-block questions went \emph{beyond} the material shown, and only 25\% could be answered from \emph{within} it.
This predominance of \emph{beyond} is what a well-formed situation model predicts: rather than re-reading the display, participants use it as a basis for inferences about mechanism, magnitude and consequence that were not explicitly stated \citep{zwaan1998situation}.
Here the arms differ only directionally, with treatment participants asking more \emph{within}-block questions (28\% vs.\ 21\%; $p=0.16$, participant-clustered).

Taken together, the findings suggest a complementary relationship between immersive information and LLM agents in civic consultation. The immersive material grounded inquiry in concrete, memorable features of the proposal, while the conversational agent allows participants to retrieve, extend, and challenge the fixed information they were given.

%% GUSTAVO
{\bf Gustavo} was engaged quite differently from Flo (\autoref{fig:chats}c,\,d).
The most common substantive theme was \highlighttag{Visual Empathy} (21\%), in which participants pictured the redesigned space as a good for others (``fruit trees can turn the area into a meeting place [\ldots] for strangers to share a fruit'').
Next came \highlighttag{Operational Barriers} (17\%), practical doubts such as ``you aren't going to be able to bring tons of fresh, cooled produce on the back of a bike,'' and \highlighttag{Civic Engagement} (15\%), reflections on consultation, communication, and legitimacy.
Together these three themes account for over half of all sentences.

Grouped by stance toward the redesign, positive themes (28\%) were as common as negative ones (23\%);
This is interesting given the real-world almost exclusively negative discussion of this project in newspaper comment sections, which we will come back to in the limitations section of this paper.
We examined Visual Empathy more closely along two axes: \emph{Strength} and \emph{origin}.
\emph{Strength} is categorized \emph{``strong''} when a beneficiary is named (``a meeting place [\ldots] for strangers to share a fruit'') versus \emph{``weak''} when only a general benefit is named (``a nicer place to be'').
\emph{Origin} indicates whether the empathic frame is \emph{participant-raised} or \emph{Gustavo-prompted}.
Interestingly, more treatment participants expressed strong \highlighttag{Visual Empathy} than control participants (63\% vs 44\%,$p=.007$).
The gap is carried entirely by participant-raised \highlighttag{Visual Empathy} (59\% vs 40\%, $p=.007$); the Gustavo-prompted variety is rare in both arms and does not separate them (8\% vs 7\%, $p=.79$).
The weak \highlighttag{Visual Empathy} does not separate the arms.
Importantly, Gustavo raises other-directed language at the same rate in both arms 
(82.4\% vs.\ 81.2\% of his turns contains at least one social or beneficiary keyword, $p=.55$), therefore any difference between arms originates on the participants' side.
Because \highlighttag{Visual Empathy} is already common in both arms, the treatment does not appear to \emph{create} other-directed imagining so much as to make it easier to render concretely: what differs is not whether participants voice a benefit for others, but whether they name who would receive it.
One reading is that the immersive walkthrough leaves participants with a richer mental model of the street: Picturing ``a meeting place [\ldots] for strangers to share a fruit'' is easier when one has a vivid spatial construct to populate, which ties the effect to the same spatial mechanism that improves recall (\autoref{fig:memory}).

%\textbf{Chat negative}
An interesting finding in the Gustavo chat was that, while there was a high prevalence of positive sentiment, there was a distinct minority of four (out of 200) participants who exhibited interaction patterns that were entirely one-sided and negative: ``My opinion is that the World Economic Forum's plan to make us all battery chickens ... Taking away... the possibility of moving where, how, and when they want ... the first step to caging them in prisons of 15 minutes and 30 km/h...''
The chats of these individuals demonstrated the limitations of the LLM: in the presence of an entrenched, extremely negative viewpoint, Gustavo was unable to encourage the consideration of positive aspects or alternative perspectives.

%specific questions
Another notable finding from both chats was the emergence of highly specific and locally grounded questions—for example: ``Are the newly planted trees well-rooted? Urban trees often fail because their roots don’t take,'' or ``Will there be services for leaf pickup and gardening?''
Such granular concerns rarely appear in standardized public consultations or official information materials, which must remain concise and cannot anticipate every issue that matters to residents. 
Yet, these questions reveal authentic concerns. A conversational LLM interface can accommodate this breadth of inquiry that would otherwise be too resource-intensive for planners to prepare in advance.

%\textbf{Answer Quality}
We also evaluated the LLMs' responses to participant queries, focusing on whether they introduced any information beyond the provided fact pack. The models were explicitly instructed to answer from that package and to offer adjacent facts and indicate when relevant information was missing. Overall, the LLMs' adherence to these constraints was high, with few exceptions. In one conversation, repeated prompting led an agent to extend beyond the fact package; in a handful of others, the agent converted a figure it did contain into the participant's own currency or units. In each case the information remained factually correct.

\paragraph{LLMs received positive ratings despite concerns about friction and bias.}

% QUANTITATIVE DATA
Perceptions of the LLM agents were generally positive (\autoref{fig:guides_rating}b, \emph{Ratings}). Participants rated the agents as most useful when thinking about other citizens: 74\% agreed or strongly agreed that the LLMs would be helpful for \highlighttag{Others}. This supports prior findings that third-person evaluations tend to be more candid than self-assessments \cite{fisher1993social}.
Self-reported perceived helpfulness was lower, but still favorable. About 60\% of participants found \highlighttag{Flo} moderately or very helpful, with similar results for \highlighttag{Gustavo}. 
\highlighttag{Trust} in the information was moderate overall, with 53\% of participants reporting very or extremely high trust.

% CRITICAL
When asked to evaluate whether the LLM agents were helpful or exerted undue influence (\autoref{fig:guides_rating}b, \emph{Open feedback}), the dominant theme was \highlighttag{Positive Utility} (64\%): ``I found them very useful and understanding, no improvement is needed.''
However, \highlighttag{System Limitations} emerged more in the control group (35\% vs. 22\%; $p=.061$), with some concerns about trustworthiness—``I don't trust them as I don't know enough about them''—and limited domain knowledge: ``They were limited in their knowledge... and could only provide certain information.''
Beyond trust issues, \highlighttag{Interaction Friction} centered on excessive questioning, verbosity, and sycophancy: ``Gustavo was continually trying to engage, which could be a bit annoying'' and ``They both suffered from a sycophantic nature, although Gustavo seemed more inclined to tell me I was insightful as often as possible.''
Participants were divided on neutrality. Some experienced \highlighttag{Reflective Neutrality}: ``I think they were neutral and didn't sway me in either direction. Gustavo encouraged me to form my own opinion.'' Others perceived \highlighttag{Bias}, viewing the AI as a ``salesperson'' for the project: ``it is aimed at always favoring the project and never really talking about the issues.''

% SEPARATE
Participants in treatment and control groups were split evenly on separation versus merging the roles (\autoref{fig:guides_rating}b, \emph{Role separation}). A prominent theme was \highlighttag{Role Utility \& Purpose}: ``Flo helps you to understand the changes, while Gustavo helps you to form your opinion on those changes.''
Yet \highlighttag{Artificiality \& Necessity} revealed frustration: ``I would have been quite happy to have just had one character to interact with, one who could cover everything in detail.''
Tensions arose between valuing \highlighttag{Clarity \& Reliability}—``Clearly separating facts from opinions is important''—and experiencing \highlighttag{Functional Issues} from handoffs: ``it doesn't feel good when someone refers you to the other person... a bit wasting my time.''
Specific critiques included Gustavo's agreement bias (\highlighttag{Gustavo Critique}): ``I find there is a tendency with LLMs to simply agree with everything you say.''
Across the open evaluations, eight participants (4\%) spontaneously remarked on the agents'---especially Gustavo's---tendency to agree, flatter, or over-reassure instead of challenging. Thus, although a minority concern, this reflects the well-documented \emph{sycophancy} failure mode of many LLMs, which stands in direct tension with the deliberative goal of surfacing counter-arguments.

\paragraph{Approval was near-universal in both arms.}
Overall, 97\% of treatment and 95\% of control participants approved the renovation project as a whole (\autoref{fig:vote}a, top row; Fisher's exact $p=0.72$).
Approval (as compared to disapprovals, and neutral) across the six sub-aspects averaged 84\% with no between-group difference on any aspect (\autoref{fig:vote}a; $\chi^2$ $p=0.38$ to $0.96$).
Ranked voting put {\em traffic} first and {\em canopy} second in both arms; the remaining four aspects fell within 0.22 rank points of one another (3.59--3.81 on a six-point scale) and did not order consistently across arms (\autoref{fig:vote}b). The single between-arm difference is {\em traffic}, ranked more important by the treatment group than by the control group (mean rank 2.53 vs.\ 3.06, Mann--Whitney $p=0.02$), which does not survive correction across the six aspects ($p=0.12$).
Notably, the different voting methods produced inconsistent results: {\em parking} drew the highest disapproval rate (10.5\%) but ranked fourth of six, while {\em sponge pavement}---frequently praised in open-ended responses---ranked last, tied exactly with {\em biodiversity}.

This is unlikely to be a memory artifact: participants' open-ended recall was largely accurate, describing the very block they were filed under in roughly 80--98\% of cases. Rather, several sub-aspects are causally entangled in the redesign itself---removing parking is what creates room for the added trees, greenery, and habitats---so {\em parking} is seldom recalled or judged in isolation: nearly two-thirds of parking recalls also invoke the green benefits it enables. Asked to rank such a bundled item on its own, participants' choices map poorly onto their integrated mental model of the street.
{\em Sponge pavement}, in contrast, was recalled as a distinct feature---only a small minority of its recalls mention another block. So, its low ranking is not an entanglement artifact, but a genuine ordering: an aspect can be appreciated in open-ended reflection, yet, weighted less when ranked against more salient priorities.
Ranking fuzzy sub-aspects of a fictitious street is also only weakly incentivised relative to the conversational tasks, which likely adds noise. For the entangled blocks, then, participants' mental representations during voting deviated from the official category definitions, which limits the interpretability of between-group comparisons at the sub-aspect level.

Overall, approval was near-universal in both arms, which leaves little room for differences to appear in the first place.
A comparable decoupling has been reported for immersive planning material, where headset display improved what citizens recalled of a design without affecting their confidence in the vote or the outcome of the ballot \cite{vanleeuwen2018effectiveness}.
What the ceiling does indicate is that this proposal was uncontroversial among our participants, and that a more divisive project would be needed to put the question to a real test.

\paragraph{Consultation proposals to the city were predominantly positive.}
After the voting and LLM dialogue phases, participants could submit open-ended feedback to the city.
Substantively, a common theme was \highlighttag{Constructive Engagement} (23\% of all themes; \autoref{tab:consultation}), covering systemic and technical dimensions of the redesign.

Another common theme was \highlighttag{Personal Concerns} (26\% of all themes), including requests for comfort infrastructure (e.g., ``installation of sun sails'') and economic and lifestyle threats (e.g., ``removal of parking spaces will destroy local business'').
The most striking pattern, however, is the overall tone. \highlighttag{Positive Sentiment} (e.g., ``I think it’s a good plan'') was the single most common theme (38\%), while outright \highlighttag{Negative Sentiment} was rare (no more than $3\%$ in either arm). Interestingly, ``treatment'' participants offered positive endorsements significantly more often than ``control'' participants ($68\%$ vs.\ $53\%$; Fisher’s Exact Test, $p=0.04$, $V=0.15$).

This constructive tenor stands in sharp contrast to the public reactions to the same real-world project in online comment sections, which were predominantly hostile (\autoref{apx:comp_real}).

\section{Conclusion and Discussion}

In this paper we conducted a human-subject experiment in the context of participatory urban planning.
We held the information content of the urban sustainability project constant and varied only the format in which it was delivered. Participants who experienced the project through 360-degree video with spatial audio, supported by the method of loci (spatial anchoring of information), recalled the information more fully than those who encountered the same facts through text and images.
The same information material also changed how participants engaged with the conversational agents. Treatment participants were more likely to name a specific beneficiary when imagining the redesigned street ($p=.007$), and directionally more likely to ask \emph{on-block} questions---about the six information blocks they had been shown ($p=0.07$). Their written feedback to the city was also more often positive ($p=0.04$; \autoref{tab:consultation}). The recall difference and these conversational differences are both consequences of the briefing; the video may have produced each of them directly, and our design does not establish that the first produced the second.

\paragraph{Information-environment recommendations for civic LLMs.}
Our findings point first to the importance of designing not only the LLM, but also the information environment in which citizens encounter it. The treatment and control conditions were designed to mirror realistic municipal communication choices. The control condition---text and images presented on a webpage---resembles the format used in most public consultations today. Likewise, the short video clips used in the treatment condition could be embedded in a consultation website and delivered to citizens in essentially the same way as in our study.

This distinction matters because the information environment shapes the conversation that follows. A consultation that relies primarily on static text and images should expect the subsequent chatbot interaction to go \emph{off-block}, away from the project's fact base: citizens may ask questions that are only loosely connected to the information they were given, or that wander entirely off the project. By contrast, a richer and more memorable information environment increases the extent to which subsequent questions stay \emph{on-block}.

For municipalities, this creates a trade-off in how the conversational agent is subsequently configured. A general-purpose model can answer almost any question, but its flexibility comes at the cost of control: it becomes harder to ensure that answers remain grounded in a predefined fact pack. A fact-pinned model, by contrast, can constrain the conversation to information that has been explicitly provided and vetted by the municipality at the cost of terser answers and of routing elsewhere the questions that fall outside that material. This was the approach taken by one of our conversational AI agents, Flo. Several participants experienced this constraint as a limitation, reporting that Flo did not really answer their questions or redirected them to Gustavo, which Flo was instructed to do whenever a question called for interpretation or fell outside the fact pack.

The implication is that municipalities should invest in the information environment that precedes and surrounds the LLM, since how a quantity is contextualized determines whether a lay audience can use it \cite{peters2007numeracy}. 
Doing so pays twice: it improves what citizens retain, which is valuable in its own right, and it increases the likelihood that the subsequent conversation remains anchored in information that the municipality can substantiate. 
The \emph{information environment therefore constitutes a first layer of civic LLM design}, before the persona or conversational behaviour of the agent itself is considered.

\paragraph{Persona recommendations for civic LLMs.}
A second layer of design concerns the persona and conversational behaviour of the LLM itself. About 60\% of participants rated the LLMs as moderately or very helpful, and half reported high levels of trust in them. Our experimental setup therefore suggests that participants experience the combination of a curated information stage and LLM-supported engagement as useful and important. At the same time, participants differed in how they evaluated the separation of agent roles. While some valued the distinction between fact- and opinion-oriented agents, others perceived the separation as redundant.

We therefore recommend retaining distinct agent roles while allowing citizens to choose freely between them. This preserves the benefits of role differentiation without imposing a conversational structure that some citizens may experience as unnecessary. Opinion-oriented agents should additionally limit follow-up questions and be explicitly tuned to challenge users' views, thereby counteracting the sycophancy that some participants criticised.
Finally, for participants exhibiting entrenched resistance to sustainability interventions, LLM-based engagement should be complemented by in-person deliberative formats.

\paragraph{A human-in-the-loop diagnostic instrument for municipalities.}
The six information blocks shown to participants reflected actual municipal communication choices emphasizing design and expected benefits, and covered neither cost nor construction timeline. The agents' shared content was broader, extending to those and other questions anticipated as common. 36\% of participants asked Flo what the project cost and 15\% how long it would take. The information was available to every participant, but reached only those who asked for it. The study thus exposed a concrete communication gap in a real municipal consultation.

The ``diagnostic capability'' of our approach depends on participants treating the consultation as consequential. 
Our experiment did not require participants to interact with either agent: they could skip the conversational stage entirely. Nevertheless, participants engaged approximately five turns with Flo on average and eight with Gustavo, and their questions remained closely anchored to the project. This level of voluntary engagement suggests that the combination of immersive presentation, mnemonic encoding, and conversational interaction did elicit substantive reactions.

We see this diagnostic function in the sense of information gap detection as a contribution in its own right. 
Pretesting how an audience will respond to a proposed intervention is valuable, and a market has consequently emerged to meet this demand. Venture-funded firms \cite{simile,electrictwin,artificialsocieties} now offer ``synthetic audiences'': LLM-simulated respondents designed to predict how a public might react to a proposal. The scientific evidence, however, warrants greater caution, as simulated audiences have not been shown to reproduce the breadth of preferences expressed by real people \cite{yang2024llmvoting}.

Our approach takes a fundamentally different, human-centered route. The criticism, questions, and ideas elicited come from \emph{human} participants. A simulated audience can suggest which questions a public might ask, such as about cost and timeline. It cannot say how many people actually needed to ask them, and that share is what tells a municipality whether an omission matters.
Moreover, we show that the diagnostic stage need not recruit the same population that will ultimately participate in the public consultation. Municipalities can therefore use human participants to stress-test communication materials without directly influencing the electorate that will later be consulted.

\paragraph{Addressing negativity bias in online sustainability discourse.}
When participants in our experiment provided feedback to the fictitious city, positive sentiment prevailed.
This contrasts sharply with comments from citizens of Lausanne, who posted in the comment sections of newspaper articles on the real project that inspired our study. Those reactions were predominantly negative and hostile (Appendix~\ref{apx:comp_real}).
The contrast of reactions in the field vs. in our experiment thus suggests that the negativity of online sustainability discourse can, at least partly, be an artifact of platform design.
Our results show that a well-designed platform can channel engagement toward substantive input, and because the resulting feedback is predominantly constructive, such consultations can transparently be displayed.

\paragraph{Limitations and generalizability.}
Several features bound the generalizability of our results. Most importantly, \emph{hypothetical bias}: participants were paid online workers (via Prolific), reasoning about a fictional city in which they had no real stake. Because the commutes, property values, and neighborhood identities that matter in real planning were only simulated, findings such as the treatment group's reduced focus on personal inconvenience---and our broader claims about real-world policy application---must be read with corresponding caution. 
Second, \emph{sample composition}: the cohort of online workers had a self-selection bias, being predominantly White (85\%), UK-based, educated, and (recruited online) being digitally comfortable (\autoref{tab:sociodemographics}).
Our inclusivity claims are therefore provisional and call for further validation with more representative populations, including digitally marginalized ones. 
Third, \emph{cross-cultural transfer}: the scenario was a European, Swiss-inspired street redesign evaluated largely by non-Swiss, English-speaking participants, such that norms around car use, parking, and public space may differ across contexts in ways that could affect responses. 
These bounds do not fall evenly across our results. The omissions the instrument exposed in the municipality's material are properties of that material and would confront any reader; the effects of presentation format on recall rest on encoding mechanisms that are not specific to a public. 
How the agents' neutrality and role separation were judged is the part that is public-dependent.
Those judgments should be read as a demonstration of what the instrument surfaces, not as an answer for the consultation it was inspired by. 
The investigated project was also relatively non-partisan: future work could test divisive topics such as housing density, renewable energy siting, or congestion pricing. Furthermore, our intervention was not effective among participants holding strongly negative prior views, for whom in-person engagement seems necessary. 
Fourth, \emph{model dependence:} both agents ran on a single pinned checkpoint, and benchmarks of LLM answers to citizen questions about government information report wide variations not only between model families but also within a single model, with median performance well above the mean and a long tail of poor answers \cite{majithia2026citizenquery}. Our reliability observations therefore describe this model when the study was performed; a different checkpoint, or the same one on a different draw, could behave differently. 
Finally, \emph{persistence:} we measure outcomes only once, immediately after the session, and therefore cannot determine whether participants' knowledge gains or reasoning about the project persist over the interval between such a session and an actual ballot. Prior work counsels caution: BallotBot's knowledge gains had disappeared when participants were re-tested one week later, both on the quiz and in the quality of their written reasoning \cite{ash2025ballotbot}. The longer-term dynamics of engagement and retention therefore remain unknown.

\paragraph{Democratic sustainability communication.}
Immersive and conversational systems require careful implementation.
While sensory-spatial features can enhance recall and engagement, depending on the kind of implementation, immersive environments can also have significant persuasive potential, but should not be used to manufacture consent \cite{fogg2003persuasive, slater2009place}.
Additionally, conversational systems should not become forms of tokenistic participation or ``window dressing,'' where citizens are invited to express opinions, without actually having a meaningful influence on decision-making processes \cite{arnstein1969ladder, sloane2022participation}.
It is, therefore, important to empirically validate and continuously assess such systems in practical applications. Importantly, in democratic contexts, the aim is not to produce unanimous consensus but to support the development of reasonable and broadly acceptable solutions, valuing diversity and pluralistic perspectives.

\section*{Acknowledgments}
We are grateful to Uzufly for providing the 3D rendering of the Lausanne city center.
The preparatory phase of this experiment was informed by the ERC project CoCi – Co-Evolving City Life.
We also thank the Decision Science Laboratory at ETH Zurich for supporting the data collection through Prolific.

\newpage
\bibliographystyle{unsrtnat} % numeric, ordered by appearance
\nocite{makasi2022typology,alishani2025citizens,mushkani2025wedesign,alsobay2026table} % printed in Fig. 1 only
\bibliography{references_chi}

\newpage
\appendix
\section{Appendix}
%\onecolumn
\setcounter{figure}{0}
\setcounter{table}{0}
\renewcommand{\thesection}{\Alph{section}}%
\renewcommand{\thesubsection}{\thesection.\arabic{subsection}}
\renewcommand\thefigure{A.\arabic{figure}}
\renewcommand\thetable{A.\arabic{table}}
\vspace{.5cm}

\begin{table*}[htb!]
\captionsetup{justification=raggedright,singlelinecheck=false}
\begin{minipage}[t]{0.51\textwidth}
\caption{Participant characteristics (N = 200).}
\label{tab:sociodemographics}
\small
\setlength{\tabcolsep}{3pt}
\begin{tabular}{l r @{\hspace{1.5em}} l r}
\toprule
\multicolumn{2}{@{}l}{\textbf{Age (years)}} & \multicolumn{2}{@{}l}{\textbf{Sex (\%)}} \\
\midrule
Mean (SD) & 40.2 (12.7) & Male & 57.0 \\
Min--Max & 20--75 & Female & 43.0 \\
25th--75th pct & 30--49 & & \\
\midrule
\multicolumn{2}{@{}l}{\textbf{Ethnicity (\%)}} & \multicolumn{2}{@{}l}{\textbf{Residence (\%)}} \\
\midrule
White & 85.0 & United Kingdom & 64.5 \\
Asian & 5.5 & Poland & 15.5 \\
Black & 4.5 & Italy & 10.5 \\
Mixed & 4.5 & Netherlands & 3.5 \\
Other & 0.5 & Germany & 3.0 \\
 & & France & 2.5 \\
 & & Belgium & 0.5 \\
\midrule
\multicolumn{4}{@{}l}{\textbf{Employment status (\%)}} \\
\midrule
Full-time & 38.0 & Part-time & 11.0 \\
Student & 18.5 & Not in paid work & 6.0 \\
Expired/missing & 22.0 & Unemployed & 3.5 \\
Starting new job & 0.5 & Other & 0.5 \\
\bottomrule
\end{tabular}
\vspace{2mm}

\footnotesize \textit{Note.} No statistically significant differences were found between the treatment and control groups for any demographic variable.
\end{minipage}
\hfill
\begin{minipage}[t]{0.445\textwidth}
\caption{Self-reported perceptions and behaviours.}
\label{tab:mobility}
\small
\setlength{\tabcolsep}{3pt}
\begin{tabular}{l r r}
\toprule
  \% & \textbf{Control} & \textbf{Treatment} \\
\midrule
\multicolumn{3}{@{}l}{\textbf{Mode of transport}} \\
Bicycle & 9.9 & 10.1 \\
Car & 46.5 & 45.5 \\
Public transportation & 21.8 & 18.2 \\
Walking & 21.8 & 26.3 \\
\midrule
\multicolumn{3}{@{}l}{\textbf{Community importance}} \\
Very important & 18.8 & 26.3 \\
Somewhat important & 48.5 & 43.4 \\
Neutral & 21.8 & 16.2 \\
Somewhat unimportant & 8.9 & 11.1 \\
Not at all important & 2.0 & 3.0 \\
\midrule
\multicolumn{3}{@{}l}{\textbf{Neighbourhood attractiveness}} \\
Very important & 71.3 & 77.8 \\
Somewhat important & 27.7 & 18.2 \\
Neutral & 1.0 & 3.0 \\
Not at all important & -- & 1.0 \\
\bottomrule
\end{tabular}
\vspace{2mm}

\footnotesize \textit{Note.} Perceived importance of transport connectivity and neighborhood attractiveness was comparable across treatment and control groups, with no statistically significant differences observed.
\end{minipage}
\end{table*}

\subsection{Experiment Ethics}
The experiment was conducted online with participants recruited via Prolific. The experimental design was approved by the ethics commission of ETH Zurich (EK 2024-N-99).
The surveys were programmed using SurveyJS. The LLMs (GPT-4o, checkpoint \texttt{gpt-4o-2024-08-06}) were accessed via the OpenAI API.
Their use was explicitly reviewed and approved by the ethics commission. Participants were repeatedly and clearly informed that they would interact with LLMs in the study. Participants were explicitly instructed not to submit personal data such as names or other identifying information in their interactions with the LLMs.
To ensure data protection, the collection of decision data was kept separate from demographic information. Sociodemographic data were downloaded from Prolific and matched via the Prolific ID but were never stored together with experimental data. The experiment ran on a server located in Switzerland to ensure compliance with data protection regulations.

\subsection{Technical Details}
\label{apx:video}
Anonymized participant data, analysis code, qualitative coding, and the complete deployed system prompts will be made available in a public GitHub repository.
The experimental stimuli were based on a high-resolution photogrammetric scan of Place de la Cathédrale in Lausanne, Switzerland.\footnote{Google Maps reference: \url{https://maps.app.goo.gl/kuN88mFuTU88rCDd9}}
The scan was acquired via drone by the professional surveying company Uzufly\footnote{\url{https://uzufly.com/}} and reconstructed as a three-dimensional digital twin.
The virtual Ville d’Ordinaire environment was developed in Unity 2022.3.36f1 with C\#, using the digital twin as the foundational asset. The environment was further modified in Blender and augmented with animated pedestrian avatars from Ready Player Me \cite{readyplayerme}, vehicle and vegetation models from the Unity Asset Store, and sound effects from Pixabay \cite{pixabay}.

\begin{figure*}[htb!]
\centering
\includegraphics[width=0.99\linewidth]{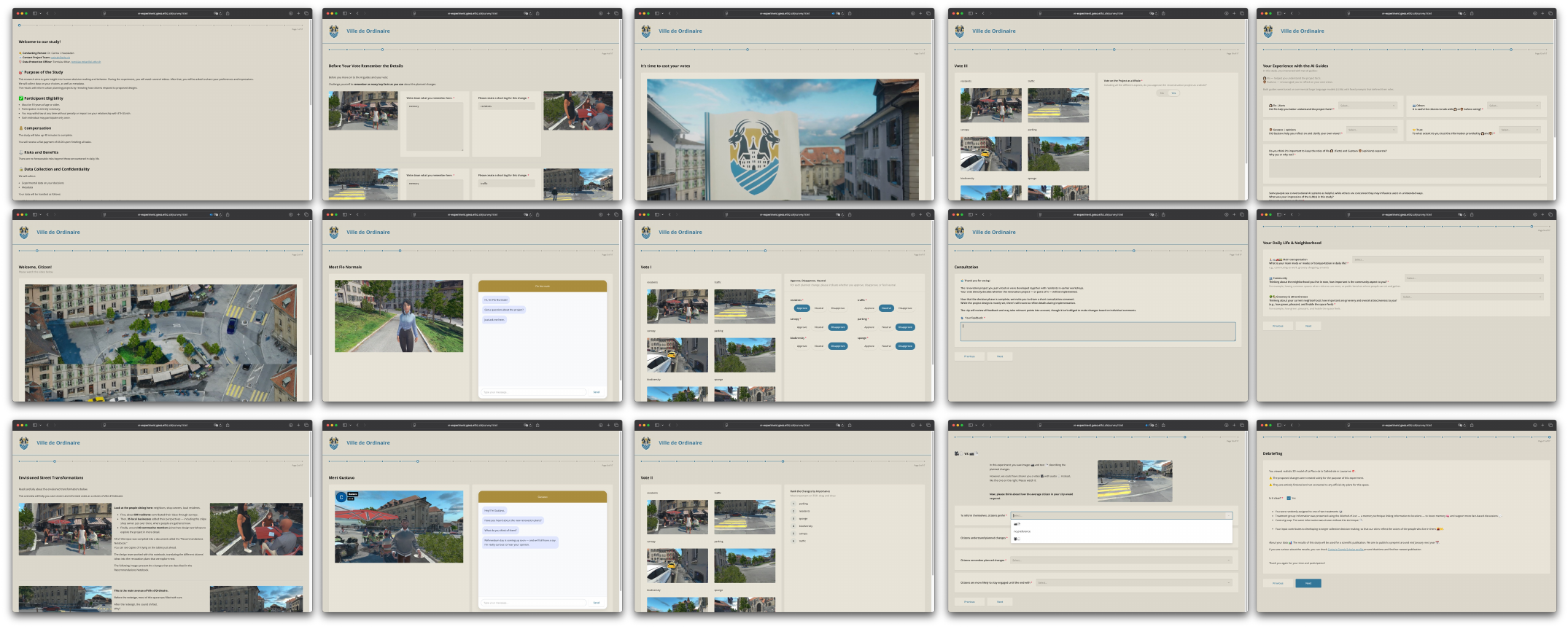}
\caption{Experimental interface and procedure (screenshots ordered column-wise). Participants completed: (1) consent, (2) introduction, (3-4) information provision—static images (control, shown) or 360-degree video (treatment), (5) memory recall, (6-7) conversations with fact-based (Flo) and deliberative (Gustavo) LLM agents, (8-10) voting tasks, (11) city consultation feedback, (12-14) evaluation of presentation format and LLMs, (15) debriefing.}
\label{fig:screenshots}
\Description{A grid of fifteen interface screenshots in column order, showing the participant's path through the study: consent, introduction, information provision as either static images or 360-degree video, the free-text recall task, the two agent chat windows, three voting tasks, the consultation feedback field, format and agent evaluation questionnaires, and debriefing.}
\end{figure*}

Video stimuli were rendered directly from the environment. Spoken narration was generated using the Chirp 3 HD voice model from Google’s Text-to-Speech service. To enhance ecological validity, the camera was positioned in a first-person perspective, and pre-recorded walkthroughs followed a natural human walking pace with subtle head-movement oscillations to simulate locomotion. Three-dimensional spatial audio was synchronized with the visual scene to further increase realism and immersion. \autoref{fig:screenshots} shows the resulting interface at each step of the participant's path through the study.

\begin{flushleft}
\footnotesize\textbf{Associated video assets: } \url{https://www.youtube.com/playlist?list=PLk7tDwnvsZsQ1K-FK5Uqm8vSPQjjVx4rY}

\end{flushleft}

\subsection{Data Details}
\begin{figure*}[htb!]
\centering
\includegraphics[width=430pt]{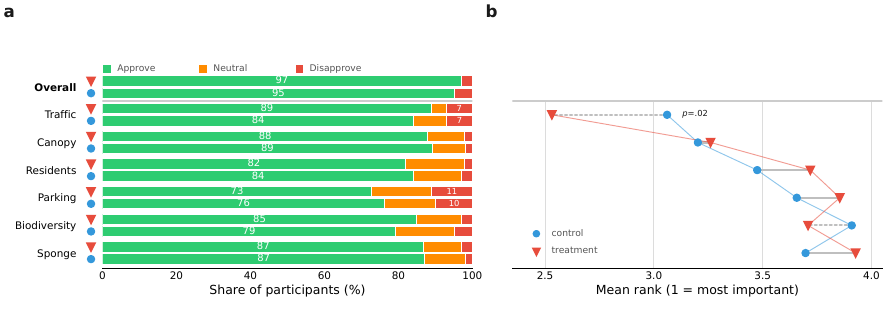}
\caption{{\bf All three ballots.} Rows are the six planning aspects, ordered by pooled mean importance rank, with the overall yes/no vote ruled off at the top. {\bf (a)} Approval voting, treatment above control within each row; percentages label the approve segment, and the disapprove segment where it exceeds 6\%. $N=200$ (101 control, 99 treatment), with no between-group difference on any row: overall Fisher's exact $p=0.72$, sub-aspects $\chi^2$ $p=0.38$--$0.96$. {\bf (b)} Mean importance rank on the same rows, control a blue circle and treatment a red triangle, the connector dotted where control ranked the aspect the more important of the two. $N=195$ (99 control, 96 treatment): the ranking was the survey's only non-required item, so the shortfall is item non-response, not exclusion. Arms are compared per aspect by Mann--Whitney $U$ and only the one contrast reaching $p<0.05$ is labelled. A ranking is compositional---six aspects competing for six slots---so the six tests are not independent, and traffic's difference does not survive Holm correction across them ($p=0.12$).
\vspace{-3pt}
}
\label{fig:vote}
\Description{Two panels sharing six rows, one per planning aspect, ordered by pooled mean importance rank with the overall yes/no vote separated at the top. Panel a shows stacked approval-voting bars, treatment above control in each row, with approval percentages labelled. Panel b plots mean importance rank per arm as paired markers joined by a connector, dotted where the control group ranked the aspect higher.}
\end{figure*}

We provide detailed tables on participants' sociodemographics from Prolific (\autoref{tab:sociodemographics}), self-reported habits in mode of transport, community importance, and neighborhood attractiveness (\autoref{tab:mobility}), as well as detailed splits in the approval voting results per information category (\autoref{fig:vote}).

\subsection{A Realistic Experimental Vignette}
\label{apx:vignette}
\phantomsection
As part of the experimental vignette, participants were exposed to a virtual scenario depicting an urban street reconstruction project. The intervention was modeled at Place de la Cathédrale in Lausanne, but its design elements were oriented at an actual, ongoing reconstruction project conducted on a different street in Lausanne (Avenue d’Echallens\footnote{\url{https://web.archive.org/web/20250909171826/https://www.lausanne.ch/vie-pratique/mobilite/espaces-publics/avenue-echallens.html}}). Key aspects of the real project—such as the creation of expanded bicycle lanes, the introduction of additional greenery, and a significant reduction in car parking—were selectively incorporated into the virtual environment.
To ensure plausibility and internal consistency, both the implementation and description of these interventions were based on planning documents published by the City of Lausanne.
Importantly, we deliberately drew inspiration from a reconstruction project on a different street from the one modeled in the digital twin. 
In addition, the virtual city was given a fictitious name (Ville d’Ordinaire) to preserve scientific independence and avoid any direct association with Lausanne. To further reduce potential bias, participants were recruited exclusively from outside Switzerland (but within a European context), ensuring that while the environment appeared recognizably European and realistic, participants were unlikely to have prior personal exposure to the actual location.

%\subsection*{Narrator Scripts (Examples)}
Narrator scripts guided participants through six information blocks in the immersive environment. Below we present three representative examples:

\textit{Block I: Residents.}
Look at the people sitting here: neighbors, shop owners, local residents. First, about 500 residents contributed their ideas through surveys. Then, 35 local businesses added their perspectives—including the crêpe shop owner just over there, where people are gathered now. Finally, around 40 community members joined two design workshops to explore the project in more detail. All of this input was compiled into a document called the ``Recommendations Notebook.'' You can see copies of it lying on the tables just ahead. The design team worked with this notebook, translating the different citizens' ideas into the renovation plans that we will explore next.

\textit{Block II: New Traffic Balance.}
This is the main avenue of Ville d'Ordinaire. Before the redesign, most of this space was filled with cars. Do you hear the steady rush of engines? Now listen again, the sound shifted. Why? Look next to the curb: A dedicated lane was painted for cyclists, to set them apart from car traffic. Also, glance at the road markings: They signal a newly reduced pace of 30 kilometers per hour. Cars can only move much more slowly now. Studies show that this speed is safer for pedestrians and cyclists to share the road with cars.

\textit{Block VI: Sponge City.}
Look down. Before the redesign, this area was a continuous layer of asphalt. Rainwater couldn't seep through; it flowed directly into pipes. Now, notice the surface beneath your feet: interlocking tiles, open in the center. About 100 square meters will be de-paved. Hear the splash ahead? Imagine it as rainfall from a summer storm. Instead of rushing straight underground, water here can filter through the tiles into a 30-centimeter layer of gravel and soil. That layer can hold and gradually release up to 2,000 liters of water for every 100 square meters—roughly ten bathtubs full of water.

\begin{table*}[htbp]
  \centering
  \caption{Consultation proposals.}
  \label{tab:consultation}
  \begin{tabular}{@{}lcccc@{}}
    \toprule
    Theme & Share & Treatment & Control & Fisher's \\
             & \%      & \%         & \%       & exact $p$ \\
    \midrule
    \multicolumn{5}{@{}l@{}}{\textit{Valence}} \\
    \quad Positive sentiment      & 37.5 & 68.0 & 53.0 & {\bf 0.04} \\
    \quad Negative sentiment      &  1.6 &  2.1 &  3.0 & 1.00 \\
    \addlinespace
    \multicolumn{5}{@{}l@{}}{\textit{Substance}} \\
    \quad Personal concerns       & 25.9 & 36.1 & 47.0 & 0.15 \\
    \quad Constructive engagement & 23.3 & 38.1 & 37.0 & 0.88 \\
    \quad Process feedback        & 11.7 & 19.6 & 18.0 & 0.86 \\
    \bottomrule
  \end{tabular}%
  \begin{flushleft}\footnotesize
  \textit{Notes.} \emph{Valence} themes record approval or rejection; \emph{substance} themes record the kind of input offered. Coding is multi-label, so a response may carry several themes (97 of 197 carry more than one) and the two groups co-occur rather than partition responses. \emph{Share} is each theme's percentage of all 317 theme assignments (the column sums to 100\%). \emph{Treatment} and \emph{Control} are the percentage of participants \emph{in that arm} who expressed the theme at least once, so those two columns need not sum to 100\%. Agreement between the human reviewer and the adjudicated LLM codes, per theme at participant level, was Cohen's $\kappa=0.81$--$1.00$ (mean $0.92$); the reviewer confirmed or corrected the LLM codes in place, so this indexes verification rather than independent double-coding.
  \end{flushleft}
\end{table*}

\subsection{Qualitative Coding}
\label{apx:qual}
We coded each qualitative variable through a three-phase, human--LLM collaborative process.
Four authors of the paper served as human coders.

\paragraph{Phase 1: Codebook development.}
The coders first independently coded the source material: for the Flo and Gustavo chats, two authors coded independently and reconciled their codes in discussion; for the remaining variables, a single author initially coded the material.
Additionally, an LLM was prompted following \cite{nguyen2025chatgpt} to discover an independent set of themes. Themes were merged where the LLM contributed relevant proposals.
Theme definitions and decision rules were then sharpened iteratively until an LLM could apply them reliably in line with the authors' reading.
This took several rounds of asking the model to justify cases the authors would have decided differently.
The resulting codebooks are explicit and thus granular.\footnote{For human annotation, agreement analysis is used to diagnose and repair ill-defined categories \cite{teruel2018increasing}. Rewriting definitions for machine comprehension therefore adds a second layer of interpretive work that renders the criteria more explicit and standardized \cite{dunivin2025scaling}. The resulting instruments are correspondingly robust: once guidelines are prescriptive, coding depends far less on prompt wording or model choice \cite{cheng2024silicon}.}

\paragraph{Phase 2: LLM-coder application and adjudication.}
The consolidated codebook was then applied by two LLM coders:
Claude Opus (checkpoint \texttt{claude-opus-4-8}) and Claude Sonnet (checkpoint \texttt{claude-sonnet-4-6}).
The LLMs coded at the sentence level, with treatment and control transcripts shuffled.
Disagreements among the two passes were adjudicated by a third LLM against the codebook to produce the final annotated dataset.

\paragraph{Phase 3: Human validation.} Two authors then independently reviewed the adjudicated LLM annotations.
The authors confirmed or corrected each tag in place against the codebook.
Where both authors agreed in disagreeing with the LLM, the authors' choice was taken;
two- and three-way disagreements were adjudicated by a human.

\paragraph{Agreement.} Agreement is Cohen's $\kappa$, computed separately per theme \cite{hallgren2012computing}
at the unit reported in each figure (participant-level presence for the recall task, turn level for the chat transcripts),
and averaged \cite{hallgren2012computing} across human--human, human1--LLM and human2--LLM agreement.
We additionally report the per-theme range \cite{artstein2008inter}.

\paragraph{Testing for treatment differences.} Where we compare arms on a code applied to chat turns, the unit of analysis is the tagged instance: a turn that raises two topics contributes one observation per topic, and a participant who writes twelve turns contributes at least twelve observations. These are not independent draws, since a talkative or idiosyncratic participant would otherwise dominate the totals. We therefore reshape each contrast into one binary row per instance and fit a logistic regression by generalised estimating equations, clustering on participant with an exchangeable working correlation, $\mathrm{logit}\,\Pr(y_{ij}=1) = \beta_0 + \beta_1\,\mathrm{treatment}_j$, so that the standard errors allow observations from the same participant to be correlated. The estimate is population-averaged, so $\exp(\beta_1)$ is the odds ratio between arms; we report the Wald $p$ on $\beta_1$.
For the on/off-block split in \autoref{fig:chats}b, for instance, 841 topic mentions from 197 participants give an odds ratio of $0.69$ on going off-block under treatment---46\% of mentions against the control arm's 55\%---with $p=0.07$. The 197 follows from take-up: 198 of the 200 participants opened a Flo chat, and one of those raised no substantive topic. Each pair of rows in that panel carries a single $p$ because the two rows are complementary---every mention is either on-block or off-block---so coding the outcome the other way round inverts the odds ratio and leaves the $p$ untouched.
Where a participant contributes exactly one observation, as in \autoref{tab:consultation}, no such correction is needed and we use Fisher's exact test.

\vspace{10pt}
\subsection{Comparison to Real Online Discussion}
\label{apx:comp_real}

The interventions and informational material were based on a real urban planning project in Lausanne, Switzerland.\footnote{\url{https://web.archive.org/web/20250909171826/https://www.lausanne.ch/vie-pratique/mobilite/espaces-publics/avenue-echallens.html}}. 
From two related articles on 24heures.ch \cite{24heures_veloboulevard, 24heures_borne_outils}, we extracted 65 comments, 64 after removing one exact duplicate; the scrape did not record authors, so the number of distinct users is not recoverable. 
The discourse was predominantly hostile, driven less by opposition to active mobility than by doubts about the redesign itself: of the 62 comments carrying a theme, the largest group concerned the infrastructure and design of the scheme (18), followed by public spending and who pays for it (11), cyclists' conduct and bike security (10), and pedestrian safety (6).

Sentiment classification for both datasets used the multilingual \highlighttag{cardiffnlp/twitter-xlm-roberta-base-sentiment} model \cite{barbieri2022}, scoring one comment or one participant turn at a time. Online comments were mostly negative: 50.0\% negative, 29.7\% positive and 20.3\% neutral ($N=64$). 
This aligns with research showing that comment sections overrepresent dissatisfied users due to self-selection and negativity bias \cite{Hu2017OnReviews, Li2008Self-SelectionReviews}, and that incivility in those sections shifts how readers perceive the issue \cite{anderson2014}.

In contrast, interactions with our LLM agents were substantially more constructive: only 13.6\% negative, compared to 35.8\% positive and 50.5\% neutral (2{,}500 participant turns from 199 participants). This shift toward positive engagement is consistent with findings that virtual interlocutors can lower the social cost of honest disclosure \cite{lucas2014}.

\subsection{LLM Implementation}
\label{apx:llm}
Both agents used OpenAI GPT-4o (\texttt{gpt-4o-2024-08-06}). Each persona was defined by a system prompt comprising (i) a role/style section unique to each agent and (ii) a shared \emph{fact pack} plus a set of guided responses to commonly asked questions. This shared content was the complete and only knowledge source available to the agents: it was supplied in full within the system prompt rather than retrieved from an open corpus, so the entire information base is fixed and auditable. We authored it from the City of Lausanne planning documentation for the Avenue d'Echallens project (\autoref{apx:vignette}). The guided-response section encodes the project's own talking points and is therefore written to acknowledge concerns and then present the project's case, including framings that treat some objections as products of habit or of opposition rather than of evidence. Some participants accordingly perceived the agents as favouring the project (see the perceived-bias results). The complete deployed prompts---role and style, fact pack, and guided responses, identical for every participant---are released verbatim in the public repository alongside the data and analysis code (\autoref{apx:video}), rather than reproduced here.

\paragraph{Generation settings.}
Both agents were queried through the OpenAI Chat Completions API with all sampling parameters left at their API defaults: temperature $1.0$, top-$p$ $1.0$, frequency and presence penalties $0$, no seed, no stop sequences, one completion per turn, and no cap on completion length (\texttt{max\_tokens} unset). 
The system prompt and the complete conversation history were resubmitted on every call, so each reply was conditioned on the entire preceding exchange. 
No retrieval, tool use, or additional moderation layer was placed on top of the API, and model output was shown to participants verbatim. 
Data were collected in October 2025.
\end{document}